\documentclass{kapedbk}
\usepackage{graphicx,amsmath,amsfonts,edbkps}
  \normallatexbib

\begin{document}
\articletitle{Interactions in high-mobility 2D electron and hole systems}

\author{E. A. Galaktionov$^1$, A. K. Savchenko$^1$, S. S. Safonov$^1$,Y. Y. Proskuryakov$^{1}$,
L. Li$^1$, M. Pepper$^2$, M. Y. Simmons$^2$, D. A. Ritchie$^2$, E. H. Linfield$^2$, and Z. D. Kvon$^3$}

\affil{$^1$ School of Physics, University of Exeter, Stocker Road,
Exeter, EX4 4QL, U.K.\\
$^2$ Cavendish laboratory, University of Cambridge, Madingley Road, Cambridge CB3 0HE, U.K. \\
$^3$ Institute of Semiconductor Physics, Novosibirsk, 630090 , Russia}

\begin{abstract}
Electron-electron interactions mediated by impurities are studied in several high-mobility two-dimensional
(electron and hole) systems where the parameter $k_BT\tau /\hbar $ changes from 0.1 to 10 ($\tau$ is the
momentum relaxation time). This range corresponds to the \textit{intermediate} and \textit{ballistic} regimes
where only a few impurities are involved in electron-electron interactions. The interaction correction to the
Drude conductivity is detected in the temperature dependence of the resistance and in the magnetoresistance
in parallel and perpendicular magnetic fields. The effects are analysed in terms of the recent theories of
electron interactions developed for the ballistic regime. It is shown that the character of the fluctuation
potential (short-range or long-range) is an important factor in the manifestation of electron-electron
interactions in high-mobility 2D systems.
\end{abstract}



\section{Introduction}

It is well known that the presence of impurities modifies electron-electron interactions in 2D
systems. The dramatic effect of disorder on electron interactions was first understood in
\cite{Altshuler,Finkelstein} -- the theory developed for the `diffusive' regime, $k_BT\tau /\hbar
\ll 1 $. This theory showed that disorder changes the electron-electron interaction rate and
introduces an interaction correction to the Drude conductivity $\sigma _0$. The conclusions of this
theory were subsequently tested in a number of experiments on low-mobility metallic systems where
the condition $k_BT\tau /\hbar \ll 1$ was realised because of small values of $\tau$. There is a
question, however, of what happens to the electron interactions in high-mobility systems where the
above condition is not satisfied, even at low temperatures. The theory of electron interactions in
the `ballistic' regime, $k_BT\tau /\hbar \gg 1 $, and in the transition between the two regimes has
recently been developed \cite{Aleiner,AleinerB} and requires experimental testing on high-mobility
2D systems.

The question of electron-electron interactions in high-mobility 2D systems is closely related to
the problem of the metal-to-insulator transition in 2D. The latter has drawn much interest after
observations  in some high-mobility 2D systems of the change in the sign of the temperature
dependence $\rho(T)$ from `metallic' (with $d\rho/dT>0$) to `insulating' (with $d\rho/dT<0$). This
change occurs with decreasing carrier density, and there is a suggestion \cite{Abrahams} that it
can be a manifestation of a critical metal-to-insulator transition which has not been expected in
2D \cite{Gangof4}. This effect is accompanied by a positive magnetoresistance in parallel magnetic
field, where a large enough field changes the character of the resistance from metallic to
insulating. In high-mobility 2D systems the change of the sign of $\rho(T)$ occurs at a low carrier
density where the interaction parameter $r_{s}=U_{C}/E_{F}\propto m^{*}/n^{1/2}$ (the ratio of the
Coulomb energy to the kinetic energy of carriers with density $n$ and effective mass $m^{*}$) is
large, so that one would expect failure of the description of these systems in terms of the
conventional Fermi-liquid approach. It is important, therefore, to understand the role of electron
interactions in the origin of the metallic behaviour and of the unusual parallel-field
magnetoresistance, as well as the applicability to the descriptions of these effects of the theory
of electron interactions in the ballistic regime -- the theory based on the Fermi-liquid
background.

After a brief discussion of the difference between the two regimes of electron-electron interactions (section
2), we present the results of our studies of electron interactions in the ballistic regime on three 2D
systems: a) a 2D hole gas in a GaAs/AlGaAs heterostructure (section 3), b) a 2D electron gas in a Si MOSFET
(section 4) and c) a 2D electron gas in GaAs/AlGaAs heterostructure (section 5). We show that the interaction
correction \cite{Aleiner,AleinerB} is present in the metallic temperature dependence of the resistance and in
the parallel-field magnetoresistance of system a). We will then demonstrate that the metallic $\rho(T)$ in
system b) can also be attributed to the interaction effect \cite{Aleiner}. An essential assumption of theory
\cite{Aleiner,AleinerB} is the point-like character of the fluctuation potential -- the assumption which is
justified in these two systems. Point-like scatterers determine the mobility in Si MOSFET structures
\cite{Ando} and very high-mobility GaAs structures with a large spacer ($d \sim 500-1000$ \AA )
\cite{Shayegan,Gold}. However, for typical high-mobility modulation-doped heterostructures with a thinner
spacer ($d < 300$ \AA) the scattering potential has a long-range character \cite{Gold,Hirakawa86} with the
correlation length equal to the spacer thickness $d$. It was shown in \cite{Gornyi,Gornyi2} that in the case
of a long-range potential the mechanism of interaction considered in \cite{Aleiner} is suppressed. Still,
electron-electron interactions can become observable if a strong perpendicular magnetic field is applied. To
examine the predictions of theory \cite{Gornyi} and compare them with the results of \cite{Aleiner} we study
the perpendicular magnetoresistance in system c): a high-mobility 2DEG in a GaAs/AlGaAs heterostructure with
$d = 200$ \AA.

The value of the interaction correction is determined by the constant $F_0^\sigma $ -- the Fermi liquid
interaction constant in the triplet channel. In the studied 2D systems $r_s$ is large: in a) $r_s=10-17$, in
b) $r_s=2-4$, and in c) $r_s=1-2.5$. In this case the value of $F_0^\sigma $ is not well known and one of the
aims of this work is to determine it experimentally for a range of $r_{s}$ and compare its value for
different systems (section 6). The value of $r_s$ in the 2DEG in Si is larger than in the 2DEG in GaAs due to
larger mass of electrons. Even larger effective mass of holes in GaAs than that of electrons in Si makes
system a) most attractive for studies of strong interaction effects. However, electrons in Si and holes in
GaAs have more complicated energy spectrum than electrons in GaAs:  two valleys in Si and a complex spectrum
of holes in GaAs. This could be the reason of the difference in some features of the effects in the 2DEG in
Si and 2DHG in GaAs (sections 3.3,4).

\section{Ballistic regime of electron-electron interaction}

The physical meaning of the condition for the diffusive regime of electron-electron interaction,
$k_BT\tau /\hbar\ll 1 $, can be understood as follows. The characteristic time required for two
interacting quasi-particles to change their energy by a value of about $k_BT$ is $\hbar/k_BT$. In
the diffusive regime this time is much larger than the momentum relaxation time $\tau$, which means
that two interacting particles are scattered \textit{many} times by impurities before their energy
exchange is realised. Electron-electron interactions in the diffusive regime give rise to the
logarithmic correction to the Drude conductivity which depends on the Fermi-liquid interaction
parameter $F$ \cite{Altshuler,Finkelstein}:
\begin{equation}
\delta\sigma_{xx}(T)=\frac{e^{2}}{2\pi^{2}\hbar}\left(1-3F/2 \right) \ln (\frac{k_BT\tau}{\hbar })  .
\label{eq1}
\end{equation}
In the ballistic regime, $k_BT\tau /\hbar\gg 1 $, the time of the energy exchange is much shorter than
$\tau$,  thus electron-electron interaction is mediated by a \textit{single} impurity. An impurity with a
\textit{short-range} scattering potential produces the modulation of the electron density (the Friedel
oscillation): $\delta\varrho\propto\frac{1}{r^{2}}\exp(i2k_Fr)$, so that an electron is backscattered from
the impurity as well as from the Friedel oscillation. The specific phase of the Friedel oscillation provides
constructive interference of the two scattered waves, which gives rise to a linear correction to the Drude
conductivity dependent on the Fermi-liquid interaction constant in the triplet channel $F_0^\sigma$
\cite{Aleiner}:
\begin{equation}
\delta\sigma_{xx}(T)=\sigma _0\left( 1+\frac{3F_0^\sigma }{1+F_0^\sigma } \right) \frac{T}{T_F}, \label{eq2}
\end{equation}
where $T_F$ is the Fermi temperature. The coefficient in the temperature dependence in Eq.
(\ref{eq2}) originates from two contributions: the first is due to exchange (Fock) processes and
the second is due to direct (Hartree) interaction. It is clear that the sign of the temperature
dependence of the conductance is determined by the sign and magnitude of the interaction parameter
$F_0^\sigma $. If $F_0^\sigma $ is a large enough negative value, $d\sigma _{xx} /dT<0$ ($d\rho
_{xx} /dT>0$) and this corresponds to metallic type of conduction.

There is another prediction of theory \cite{Aleiner}: for a wide range of values of parameter $F_0^\sigma $
the model allows a change in the sign of $d\rho_{xx} /dT$ with parallel magnetic field from positive to
negative -- the effect seen in recent experiments on high-mobility 2D systems \cite{Abrahams}. Magnetic field
suppresses the triplet term in Eq. (\ref{eq2}), so that the sign of $d\sigma _{xx} /dT$ becomes positive. It
is important to note that the resulting positive correction to the Drude conductivity in magnetic field is
expected to be universal and independent of the interaction parameter $F_0^\sigma $:
\begin{equation}
\delta \sigma_{xx}(T) =\sigma _0^B\frac T{T_F} \mathrm{\quad at\quad } B\geq B_s.  \label{eq3}
\end{equation}
Here $B_s$ is the field of full spin polarisation of the 2D system, $B_s=2E_F/\left( g^{*}\mu _B\right) $
(where $g^{*}$ is the Lande g-factor, $\mu _B$ is the Bohr magneton), and $\sigma _0^B$ is the Drude
conductivity in magnetic field.

Model \cite{AleinerB} considers the parallel-field magnetoresistance at a given finite temperature. At
small fields this theory gives a simple prediction in the ballistic regime for the magnetoconductivity
$\Delta \sigma =\sigma (B_{||},T)-\sigma (0,T)$.  The analytical expression for fields
$x=\frac{Ez}{2k_BT}\leq 1+F_0^\sigma $ (provided $-0.45\leq F_0^\sigma \leq -0.25$), is approximated
with 2\% accuracy by:
\begin{equation}
\Delta \sigma (B_{||})=\frac{2F_0^\sigma }{1+F_0^\sigma }\sigma _0\frac T{T_F }K_b\left(x,F_0^\sigma \right)
, \label{eq4}
\end{equation}
where $E_z=g_{o}\mu _BB_{||}$, $g_{o}$ is the bare \textit{g}-factor (without taking into account the
renormalisation of the $g$-factor due to interactions), and $K_b\left( x,F_0^\sigma \right) \approx
x^2f(F_0^\sigma )/3$, $f\left( z\right) =1-\frac z{1+z}\left[ \frac 12+\frac 1{1+2z}- \frac
2{(1+2z)^2}+\frac{2\ln (2(1+z))}{(1+2z)^3}\right] $. (At larger fields, $x\gg 1$, the theory predicts a
linear magnetoresistance in the range of the fields where the model is applicable:
$E_z\leq(1+F_0^\sigma)^2E_F$ .)

The theory of electron interaction in the ballistic regime in the case of a \textit{long-range} scattering
potential is developed in \cite{Gornyi,Gornyi2}. In this case the correction discussed in
\cite{Aleiner,AleinerB} is expected to be exponentially small, because for a long-range potential both the
Friedel oscillation and the required backscattering are weak. However, it was demonstrated in
\cite{Gornyi,Gornyi2} that applying a classically strong perpendicular magnetic field increases the
probability of an electron to be scattered back and this restores the electron-electron interaction. Theory
\cite{Gornyi} shows that in strong \textit{perpendicular} magnetic fields, $\omega_c \tau>1$ ($\omega _c$ is
the cyclotron frequency), electron interactions in the ballistic regime result in a parabolic, temperature
dependent negative magnetoresistance. Its magnitude is determined by the interaction correction $\delta
\sigma _{xx}\left( T\right) $:
\begin{equation}
\rho _{xx}=\frac 1{\sigma _0} + \frac 1{\sigma _0^2}\left(\mu ^2B^2\right)\delta \sigma _{xx}\left( T\right)
, \label{eq5}
\end{equation}
where $\mu =e\tau/m^*$. The method of determining the interaction correction $\delta\sigma_{xx}(T)$ from the
perpendicular magnetoresistance was originally used in the diffusive regime \cite{Paalanen}. A relation
similar to Eq. (\ref {eq5})  was derived there using the fact that in the diffusive regime of interactions
$\delta \sigma _{xy}\left(T\right)=0$ \cite{Altshuler}. As in the diffusive regime strong magnetic fields do
not change the interaction correction $\delta \sigma _{xx}(T)$ \cite{Houghton}  (provided the effect of
Zeeman splitting on interactions \cite{Altshuler} is negligible), the analysis of the magnetoresistance gave
the authors of \cite{Paalanen} the interaction correction to the conductivity  $\delta \sigma _{xx}\left(
T\right)$ at $B=0$.  In the ballistic regime, however, $\delta \sigma _{xx}(T)$ found from the negative
magnetoresistance in Eq. (\ref{eq5}) will be significantly different from that in Eq. (\ref{eq2}) derived for
$B=0$ for a short-range potential. The ballistic-regime theory \cite{Gornyi} predicts a specific temperature
dependence of the interaction correction in strong fields $\delta \sigma _{xx}\left( T\right)$ which we will
compare with experimental results in section 5.

\section{Interaction effects in a 2D hole gas in GaAs}

\subsection{Samples}

Experiments have been performed on a 2DHG in a $\left( 311\right) A$ modulation doped GaAs/AlGaAs
heterostructure with spacer $d=500$ \AA \, and a maximum mobility of $6.5\times 10^5$ cm$^2$V$^{-1}$s$^{-1}$.
This system shows the crossover from metal to insulator at $p\sim 1.5\times 10^{10}$cm$^{-2}$
\cite{Proskuryakov,ProskuryakovInt}. The hole density $p$ in the metallic region is varied in the range
$(2.09-9.4)\times 10^{10}$cm$^{-2}$ which corresponds to a value of the interaction parameter $r_s=10-17$.
(The latter was calculated using the value of the effective mass $m^*=0.38m_0$.)

It is expected that at large values of $r_s$ interactions renormalise the effective mass, which has to
increase with decreasing carrier density (this effect has been seen in 2DEGs in Si
\cite{PudalovPRL,Shashkin}). We have analysed the temperature dependence of Shubnikov--de Haas (SdH)
oscillations measured in weak magnetic fields and extracted the value of the effective mass at different hole
densities from $2.9\times 10^{10}$ to $8.2\times 10^{10}$ cm$^{-2}$ (close to the boundaries of the studied
range). The results show \cite{ProskuryakovA} that in spite of large $r_s$ the effective mass does not show
any significant density dependence: $m^*=(0.38\pm 0.02)m_0$. Its value at the lowest density $p=2.9\times
10^{10}$cm$^{-2}$ is in good agreement with the value $m^{*}=(0.37-0.38)m_0$ previously reported for hole
densities above $7\times 10^{10}$ cm$^{-2}$ \cite{Stormer}.

\begin{figure}[hbtp]
\centering
\includegraphics*[width=0.6\textwidth] {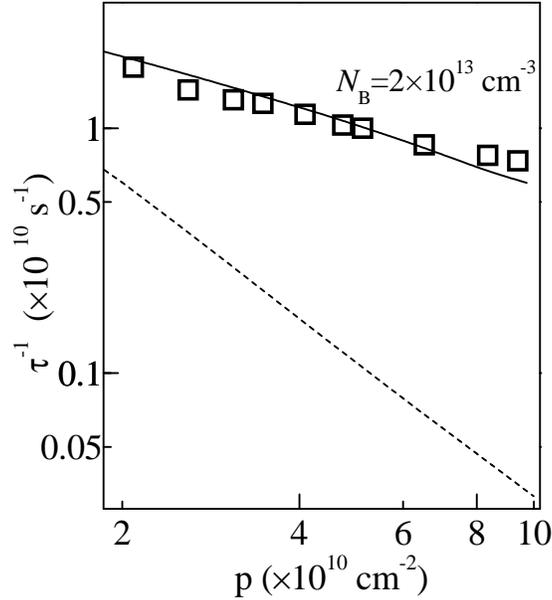}
\caption{Density dependence of the momentum relaxation rate. Symbols -- experiment, solid line -- calculation
for homogeneous background doping, dashed line -- calculation for remote doping.\label{One}}
\end{figure}

To establish the character of the fluctuation potential experienced by the holes in our structure we have
calculated the expected momentum relaxation rate $\tau^{-1}(T=0)$ at different hole densities for both
homogeneous-background and remote-doping scattering. To do this we use the experimental parameters of the
studied structure (the spacer thickness and doping concentration) and the expressions in \cite{Gold,VanHall}
for $\tau^{-1}$ in terms of these parameters. In Fig. \ref{One} the results of the calculations are plotted
together with the experimental values obtained from the relation $\rho _0=m^{*}/\left( e^2\tau p\right) $ ,
where $\rho _0$ is the value of the resistivity measured at lowest temperatures of experiment in the metallic
regime $ p \geq 2\times 10^{10}$ cm$^{-2}$.
As seen in the plot, the values of $\tau ^{-1}(p,T=0)$ calculated for remote-acceptor scattering (dashed
line) are an order of magnitude smaller than the experimental ones.

The calculated result for homogeneous-background doping is depicted by the solid line. To plot it we use the
density of the background impurities $N_B$ as the only adjustable parameter.  The obtained value $N_B=2\times
10^{13}$ cm$^{-3}$ is close to the value expected for the wafer growth conditions: $(3-5)\times 10^{13}$
cm$^{-3}$. This is also close to typical values for \emph{n}-type heterostructures with $\mu \sim 3\times
10^5-10^7$ cm$^2$V$^{-1}$s$^{-1}$ and a comparable spacer width 300-700 \AA \,: $N_B\sim 10^{13}-10^{14}$ cm$
^{-3} $ \cite{Shayegan,Gold}. Thus, one can conclude that the dominant scattering in our system is due to
background impurities with a \textit{short-range} random potential. This means that the assumption of theory
\cite{Aleiner,AleinerB} is indeed applicable to our 2DHG structure.

\subsection{Temperature dependence of the conductance and the origin of its metallic behaviour}

\begin{figure}[hbtp]
\centering
\includegraphics*[width=0.99\textwidth] {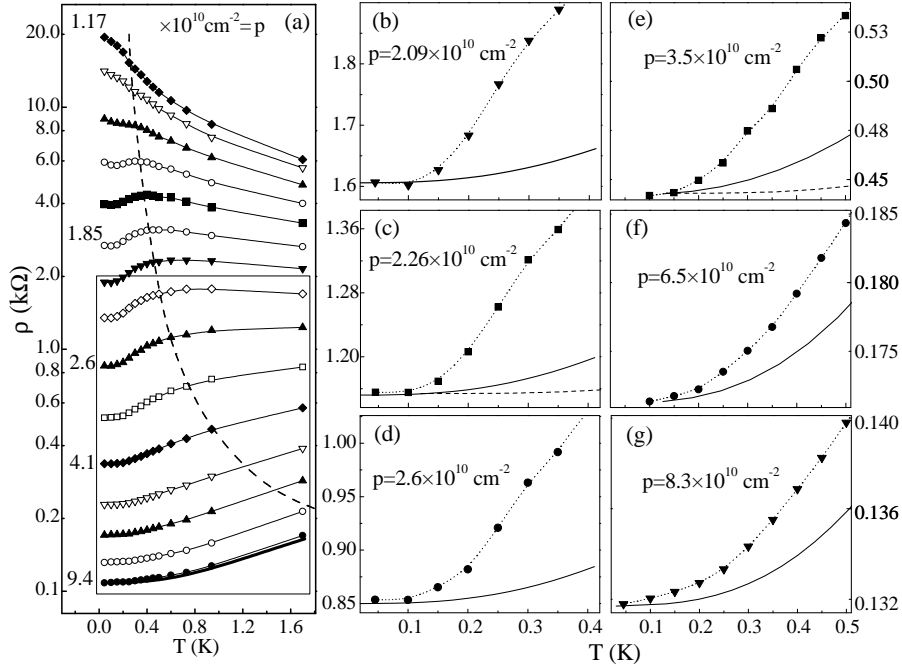}
\caption{(a) Temperature dependence of the resistivity at different hole densities near the change in the
sign of $d\rho /dT$. Bold line at the bottom of the plot ($p=9.4\times 10^{10}$cm$^{-2}$) is the calculated
$\rho (T)$ due to phonon scattering. Dashed line marks $0.3T_F$ temperature. (b-g) Resistivity on the
metallic side (symbols), together with the calculated contribution to $\rho (T)$ due to phonon scattering
($m^*=0.38m_0$). Dashed line - the phonon contribution from \cite{Noh}. \label{Two}}
\end{figure}

Fig. \ref{Two}(a) presents the temperature dependence of the resistivity, with the box indicating the curves
analysed in this work. In principle, an increase of the resistivity with $T$ can be simply due to phonon
scattering which is stronger in GaAs than in Si due to additional piezo-electric coupling.  Using
calculations for GaAs in \cite{Karpus} we have estimated the contribution of the phonon scattering as $\rho
_{ph}(T)=\frac{a(T/T_0)^3}{1+c(T/T_0)^2}$, where parameters $a$ and $c$ depend on the carrier density,
effective mass and crystal properties, and $T_0=k_B^{-1}\sqrt{2m^{*}S_t^2E_F}$, where $S_t$ is the transverse
sound velocity. The results of these calculations are shown in Fig. \ref{Two}(b-g). One can see that at
densities $p\leq 4\times 10^{10}$ cm$^{-2}$ the phonon contribution can be neglected, although at higher
densities it should be taken into account. To extend the range of the analysed data we have to subtract the
phonon contribution at $p>4\times 10^{10}$ cm$^{-2}$. The obtained results for $F_0^\sigma \left( p\right) $
agree with the trend seen at lower densities, which justifies the chosen procedure. (In a recent paper
\cite{Noh} there is a claim that our estimation of the phonon contribution is exaggerated. Fig.
\ref{Two}(c,e) shows the calculation of phonon contribution from \cite{Noh} for two hole densities close to
those in this work. Using these calculations we obtain the difference in the value $F_0^\sigma$ for these
densities less then 7\% compared with our analysis in \cite{ProskuryakovInt,ProskuryakovA}.)

\begin{figure}[hbtp]
\centering
\includegraphics*[width=0.95\textwidth] {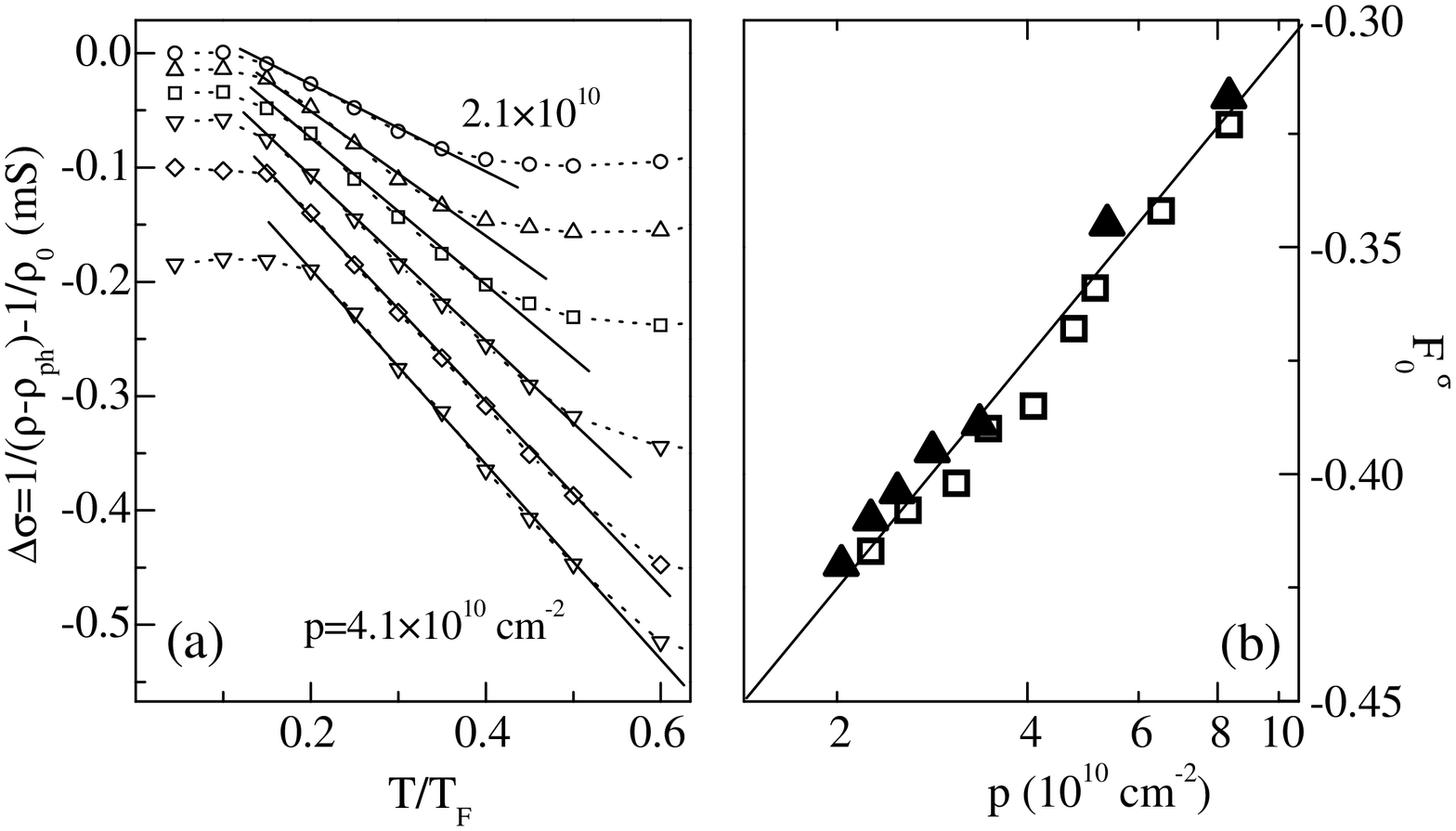}
\caption{(a) Conductivity as a function of dimensionless temperature at different hole densities, with linear
fitting. (For clarity, curves are offset vertically from the zero value at $T=0$.) (b) Fermi liquid parameter
versus hole density. Open symbols show the result obtained from the analysis of $\rho (T)$ at zero magnetic
field; closed symbols show the result from the analysis of the parallel-field magnetoresistance, Fig.
\ref{Six} (b).\label{Three}}
\end{figure}

Fig. \ref{Three}(a) shows the metallic temperature dependence of the conductivity replotted as a function of
$T/T_F$ after subtracting the residual resistivity due to impurity scattering $\rho _0=\rho \left(
T=0\right)$ (obtained by extrapolation of experimental curves to $T=0$). The peak in $\rho (T)$ in
Fig.\ref{Two}(a), with a maximum at $T_{\max }\approx 0.3T_F$ (a dip in $\Delta \sigma \left( T\right) $), is
in qualitative agreement with the expectation that after the transition to the non-degenerate state the
resistance should decrease with increasing temperature \cite{DasSarmaHwa,Mills}.

In addition to the phonon scattering, there is another possible explanation of the metallic character of the
temperature dependence. According to \cite{Murzin,Sivan}, $d\rho /dT>0$ in a high-density 2DHG in GaAs
structures can be due to inelastic scattering between two subbands which are split due to strong spin-orbit
interactions. The metallic behavior is then accompanied by positive magnetoresistance (PMR) in a magnetic
field perpendicular to the plane. As shown in \cite{ProskuryakovA}, at the highest studied hole density
$p=9.4 \times 10^{10}$cm$^{-2}$, a weak PMR of a similar shape to that in \cite{Sivan} is observed. However,
it is shown that the effect of the band-splitting on the increase of the resistivity with increasing
temperature at this density cannot exceed $3\%$. This is negligible in comparison with the experimental
resistivity increase of about $50\%$. At lower densities this effect becomes even weaker. This conclusion
agrees with the result of \cite{Yaish01}, where the band splitting is only seen at $p>1.36 \times10^{11}$
cm$^{-2}$.

After eliminating all other possibilities, we apply to the metallic $\rho (T)$ the theory of electron
interactions \cite{Aleiner}. In order to compare the results in the low-temperature range of $\rho (T)$ with
Eq. (\ref{eq2}), in Fig. \ref{Three}(a) we plot the data in the conductivity form: $\Delta \sigma (T)=\rho
(T)^{-1}-\rho _0^{-1}$. The condition for the ballistic regime $k_BT\tau /\hbar \geq 1$ is satisfied in our
structure at $T>50-100$ mK, and a linear fit of $\Delta \sigma (T)$ gives the value of the parameter
$F_0^\sigma $, Fig. \ref{Three}(b). (It should be noted that the accuracy in determining $m^*$ does not
affect our results \cite{ProskuryakovA}.)

\subsection{Magnetoresistance in parallel magnetic field}

\begin{figure}[htbp]
\centering
\includegraphics*[width=0.95\textwidth] {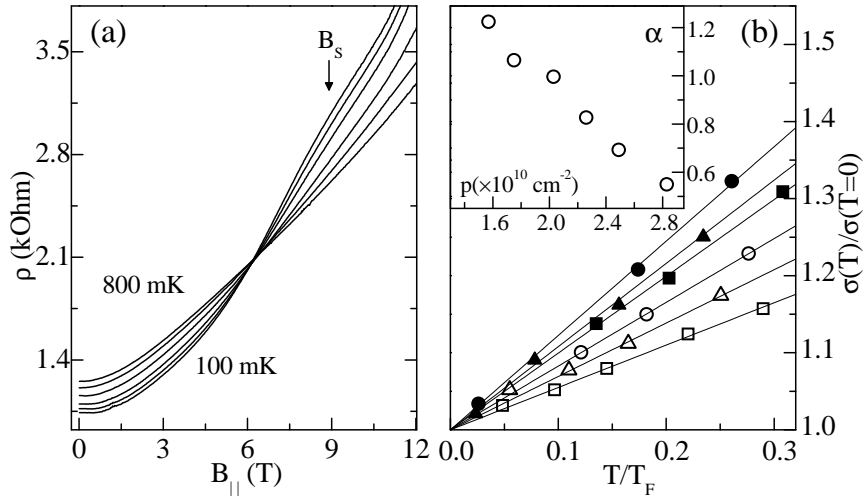}
\caption{(a) $\rho (B_{||})$ for $p=2.49\times $ 10$^{10}$ cm$^{-2}$ at different temperatures: $T=0.1$,
$0.2$, $0.3$, $0.45$, $0.6$, $0.8$ K. (b) Temperature dependence of the conductivity at $B_{||}=B_s$ for
different hole densities. Inset: The value of the coefficient $\alpha$ in the temperature dependence $\delta
\sigma_{xx}(T) =\alpha\sigma _0^BT/T_F$ for different densities.} \label{Four}
\end{figure}

Figure \ref{Four}(a) shows parallel-field magnetoresistance at different temperatures. It is seen that the
magnetoresistance is always positive. It is also seen from the temperature dependence of the resistance at
different field that $B_{||}$ drives the metallic state into the insulator at $B>$6 T. It was earlier shown
that the hump in $\rho (B_{||})$ corresponds to the magnetic field $B_s$ of full spin polarisation of the
2DHG \cite{gTutuc}. To compare the temperature dependence of the resistance in this case with the prediction
given by Eq. (\ref {eq3}) we analyse the temperature dependence of the resistivity for several densities at
several fields $B\geq B_s $. The resulting dependences, Fig. \ref{Four}(b), are indeed linear.  By
extrapolation to $T=0$ we find the value of the Drude conductivity $\sigma _0^B$, and determine the slope
$\alpha $ of the straight lines in the temperature dependence $\delta \sigma_{xx}(T) =\alpha\sigma
_0^BT/T_F$. Its value is close to the expected $\alpha =1$, although agreement is better for smaller $p$. The
smaller value of $\alpha$ at larger densities can be attributed to the fact that in a real system the
scatterers are not exactly point-like, but with decreasing density and increasing Fermi wavelength, $\lambda
_F\propto p^{-1/2}$, the approximation of short-range scatterers becomes more applicable. The observation
that at fields larger than the polarisation field the temperature dependence of the conductance is linear and
is close to the expected universal behaviour gives a strong support to the applicability of the interaction
theory \cite{Aleiner}.
\begin{figure}[htbp]
\centering
\includegraphics*[width=0.95\textwidth] {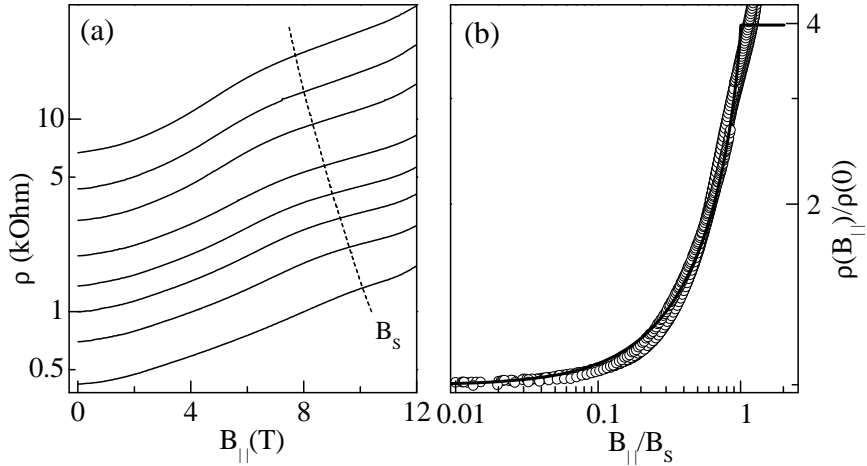}
\caption{(a) Dependence of the resistivity on parallel magnetic field at $T=50$ mK and $p=$1.43; 1.57; 1.75;
2.03; 2.26; 2.49; 2.83; $3.36\times 10^{10}$ cm$^{-2}$, from top to bottom. (b) Scaled data, with an added
curve $\rho (B_{||})$ for $p=8.34\times $10$^{10}$ cm$^{-2}$; solid line is the result of the model
\cite{Dolgopolov-Gold}. \label{Five}}
\end{figure}

From the position of the hump in Fig. \ref{Five}(a) one can determine the polarisation field $B_s$ for
different hole densities and, assuming that the effective mass is density independent \cite{ProskuryakovA},
then find the density dependence of the effective $g$-factor using the relation
$g^*\mu_B=2E_F=2p\pi\hbar^2/m^*$. As the position of the hump is not well defined, to increase the accuracy
we use a different method of determining $B_s$. The perturbation model \cite{AleinerB} can only describe a
small variation of the resistance with parallel magnetic field (see later).  There is another model which
considers the effect of a parallel field on impurity scattering \cite{Dolgopolov-Gold}. Its limitation is
that it is only valid at $T=0$, although it describes the variation of the resistance up to the field of full
spin polarisation where it predicts an increase of the resistance by a factor of four and then a saturation
of the magnetoresistance at $B \geq B_s$. This is very close to the experimental change of the resistance up
to the hump in the $\rho(B)$-curve, although there is no saturation at fields $B\geq B_s$. (We suggest that
the absence of the saturation is caused by a contribution from another mechanism  at larger fields
\cite{DasSarma-parB} -- the effect of parallel field on the orbital motion in a 2D system with finite
thickness.) An essential feature of the model \cite{Dolgopolov-Gold} is that the magnetoresistance is
dependent on the ratio $B/B_s$. In Fig. \ref{Five}(b) we plot $\rho (B_{||})/\rho (B_{\Vert }=0)$ at the
lowest experimental temperature as a function of dimensionless magnetic field $B/B_s$, with $B_s$ found as a
fitting parameter. Its value does indeed correspond to the position of the hump, Fig. \ref{Five}(a). In
accordance with \cite {Dolgopolov-Gold} all the data collapse onto one curve which is close to the
theoretical dependence, apart from the region near $B_s$.

\begin{figure}[htbp]
\centering
\includegraphics*[width=0.95\textwidth] {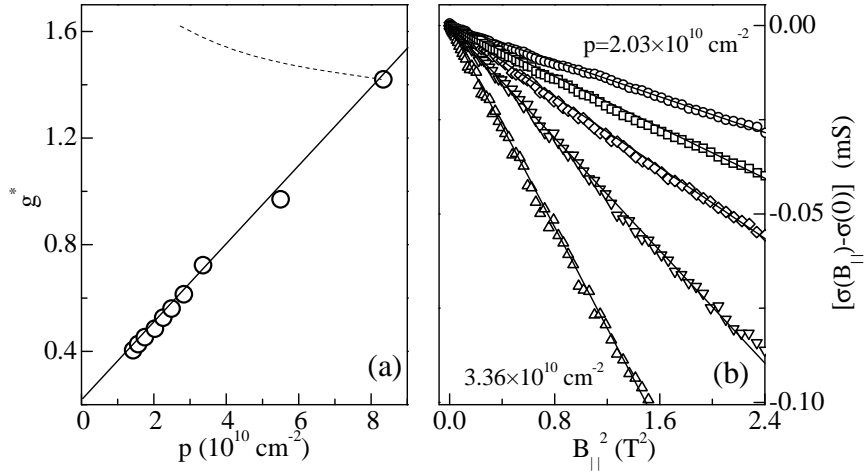}
\caption{(a) Dependence of the effective $g$-factor on the hole density, obtained from the value of $B_s$.
Dashed line shows expected renormalisation of the $g$-factor according to experimental values $F_0^\sigma
\left( p\right) $ from Fig. \ref{Three}(b). (b) Magnetoconductivity against $B_{||}^2$ at $T=0.6$ K for
densities $ p=2.03$, $2.26$ , $2.49$, $2.83$, $3.36\times 10^{10}$ cm$^{-2} $.\label{Six}}
\end{figure}

Fig.\ref{Six}(a) shows the obtained value of the effective $g$-factor, $g^{*}=2E_F/\left( \mu _BB_s\right) $,
as a function of the hole density. It is seen that the $g$-factor of holes in GaAs decreases with decreasing
density -- a similar behaviour was recently observed for 2D electrons in GaAs \cite{mTutuc}. This behaviour
of the $g$-factor does not agree with the expectation of the interaction theory, where the renormalisation of
the bare $g$-factor should lead to an increase of $g^*$:
\begin{equation}
g^{*}=\frac{g_0}{1+F_{0}^{\sigma}}. \label{eq6}
\end{equation}
Such an increase was recently seen in the 2DEG in Si \cite{PudalovPRL,ShashkinPRL}. Taking into account the
obtained values $F_{0}^{\sigma}$ in Fig. \ref{Three}(b), one would expect that interactions should give rise
to an increase of $g^*$ by about 15\% with decreasing density down to $2\times 10^{10}$ cm$^{-2}$. Instead, a
\textit{decrease} of more than three times is observed.

This decrease of the $g$-factor can be attributed to the complex band structure of holes in GaAs. It is
expected that in a 2D hole system the bare $g$-factor measured in parallel magnetic field is close to zero
when $k_{||}$ approaches zero \cite{vanKesteren}. Therefore, we suggest that the decrease of the $g$-factor
with lowering $p$ and approaching $k_{||}=0$ is caused by the behaviour of the \textit{bare} $g$-factor. It
is interesting to note that the decrease of the $g$-factor with lowering the carrier density was recently
observed for 2D holes \cite{Noh} and electrons \cite{mTutuc} in GaAs. Recently, it was suggested in
\cite{Tutuc2} that for a 2DEG in GaAs such $g\left( n\right) $-dependence can be caused by the effect of
parallel field on the effective mass due to the finite thickness of the electron channel. It was also
proposed in \cite{Zhu} that the value of the $g$-factor of the 2DEG in GaAs varies with increasing parallel
magnetic field which can account for its unusual behaviour \cite{mTutuc}. We believe that in our case of the
2DHG the complexity of the energy spectrum of holes is the dominant factor, although the effect of the other
factors requires further investigation.

Detailed analysis of the magnetoresistance at finite temperature is done at small fields where the
interaction theory \cite{AleinerB} gives a simple prediction for the magnetoconductivity $\Delta \sigma
=\sigma (B_{||},T)-\sigma (0,T)$ in the ballistic regime, Eq. (\ref{eq4}). To realise the low-field
condition, we analyse the magnetoresistance data at high temperatures. In Fig. \ref{Six}(b) we plot the
magneto-conductivity at $T=0.6$ K as a function of $B_{||}^2$ for fields satisfying the condition for Eq.
(\ref{eq4}). We use $\sigma _0$ obtained in the above analysis at $B_{||}=0$. Instead of $g_{o}$ we use the
value of $g^{*}$ determined from the analysis of $\rho (B_{||})$ at the lowest $T$, Fig. \ref{Six}(a). After
that the only unknown parameter in the slope of $\Delta \sigma (B_{||}^2)$ is $F_0^\sigma $. We extract its
value and compare it in Fig. \ref{Three}(b) with that determined earlier from the temperature dependence
$\rho (T)$ at zero field.

One can see that good agreement between two methods of determining $F_0^\sigma(p)$ is achieved, which proves
the validity of the interpretation of the results in terms of the interaction theory. The results show that
we have a good description of the metallic $\rho (T)$ and parallel-field magnetoresistance in small field by
the interaction theory in the ballistic regime.  The obtained values of $F_0^\sigma$ agree with the
description of the system in terms of the Fermi-liquid theory, in spite of the large values of $r_s$:
$r_s=17$. Indeed, according to Eq.(\ref{eq6}) the expected ferromagnetic (Stoner) instability should occur at
$F_0^\sigma=-1$, while the largest negative value in experiment is $-0.42$. The change from metallic to
insulating $\rho (T)$ occurs at the hole density $1.5\times 10^{10}$ cm$^{-2}$, where extrapolation of the
obtained dependence $F_0^\sigma(p)$ gives a value much smaller in magnitude than that required for the Stoner
instability.

\section{Electron-electron interaction in the ballistic regime in a 2DEG in Si}

The vicinal samples are high-mobility n-Si MOSFETs fabricated on a surface which is tilted from the $\left(
100\right) $ surface around the $\left[ 011\right] $ direction by an angle of $9.5^{\circ }$. The studied
samples have a peak mobility of $2\times 10^4$ cm$^2$V$^{-1}$s$^{-1}$ at $T=4.2$ K. The electron density has
been varied in the range $2\times 10^{11}-1.4\times 10^{12}$ cm$^{-2}$.

\begin{figure}[htbp]
\begin{center}
\includegraphics[width=0.95\textwidth]{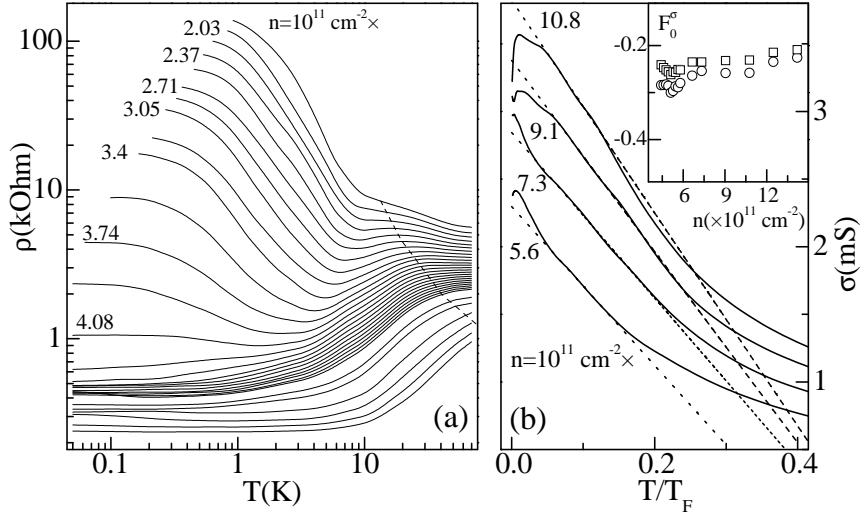}
\caption{(a) Temperature dependence of the resistivity at different densities for the 2DEG in the vicinal Si
MOSFET. Dashed line marks the Fermi temperature $T_F$. (b) An example of a linear fit of the same data in the
conductivity form, for four densities. The inset shows the parameter $F_0^\sigma $ as a function of the
electron density. The upper set of data is obtained using the dependence $m^*\left( n\right) $ from
\cite{Pudalov}. \label{Seven}}
\end{center}
\end{figure}

The temperature dependence of the resistivity has been measured in a wide temperature range, Fig.
\ref{Seven}(a). It is seen that the dependence changes with decreasing $n$ from metallic to insulating,
although, in general, $\rho (T)$ has a complicated non-monotonic character. The low-temperature results were
analysed in detail in \cite{Safonov}. Here we concentrate only on the metallic behaviour seen at larger
densities at $T>4$ K and before the transition to the nondegenerate state at $T\sim T_F$ (marked by a dashed
line in Fig. \ref{Seven}(a)). The phonon scattering can be neglected in this regime as in Si structures it
only becomes important at $T>100$ K \cite{Ando}.

The presence of the two valleys in Si modifies the expression for the temperature dependence of the
interaction correction Eq.(\ref{eq2}), which will now depend on the intensity of the intervalley scattering
and the ratio of the valley splitting $\Delta$ to $k_BT$. For weak intervalley scattering and $\Delta\ll
k_BT$ the increased degeneracy of the system due to the presence of the two valleys modifies the triplet
term, giving rise to the following temperature dependence:
\begin{equation}
\delta\sigma_{xx}(T)=\sigma _0\left( 1+\frac{15F_0^\sigma }{1+F_0^\sigma } \right) \frac{T}{T_F}. \label{eq7}
\end{equation}
(If $\Delta> k_BT$ the Hartree term becomes equal to $\frac{7F_0^\sigma }{1+F_0^\sigma}$, and for $\Delta>
E_F$ it is reduced to the conventional triplet term $\frac{3F_0^\sigma }{1+F_0^\sigma}$ in Eq.(\ref{eq2})
\cite{Vitkalov,PrivateNarozhny}. Intensive intervalley scattering will also result in Eq.(\ref{eq2})).

Fig. \ref{Seven}(b) shows an example of the temperature dependence in the conductivity form. We fit the
linear part of $\sigma (T)$ by the theory of interactions in the ballistic regime, using Eq.(\ref{eq7}). The
obtained parameter $F_0^{\sigma}$ increases with decreasing density, Fig. \ref{Seven}(b, inset), in agreement
with previous results on the 2DHG, Fig.\ref{Three}(c). We have also performed the analysis of the linear
$\rho \left( T\right) $ using Eq.(\ref{eq2}). We will later compare the values of the interaction parameter
$F_0^\sigma$ obtained with and without assumption of the valley degeneracy (section 6).

In the above analysis we have ignored the possibility of an increase of the effective electron mass with
decreasing density, taking into account our range of not very large $r_s$. Using the results of
\cite{Pudalov} where such an increase was reported for a 2DEG on $(100)$ Si we show in Fig.\ref{Seven}(b,
inset) that the effect of $m^*\left( n\right) $ could only result in an error of less than 10\%.

\section{Interaction effects in the ballistic regime in a 2DEG in GaAs. Long-range fluctuation potential.}

To examine the prediction of interaction theory for a long-range scattering potential \cite{Gornyi} we have
used a 2DEG in a standard modulation doped n-GaAs heterostructure with a thin spacer $d=200$ \AA. The
mobility changes in the range $0.42-5.5\times10^{5}$cm$^{2}$V$^{-1}$s$^{-1}$ when the electron density is
increased from $0.46\times10^{11}$ to $2\times10^{11}$ cm$^{-2}$. This allows us to vary the parameter
$k_{B}T\tau/\hbar$ in a broad range from 0.04 to 3.8 in the studied temperature interval $T=0.2-1.2$ K.

\begin{figure}[htbp]
\begin{center}
\includegraphics[width=0.95\textwidth]{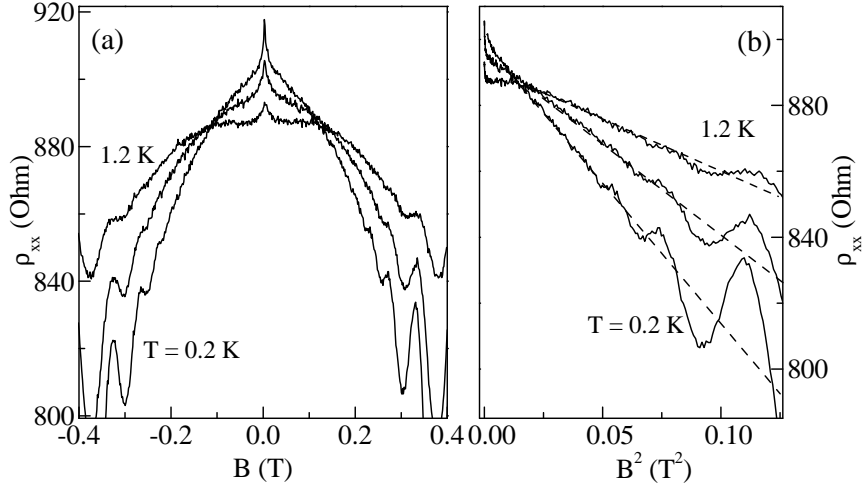}
\caption{(a) Longitudinal resistivity versus magnetic field for electron density
$n=6.8\times10^{10}$cm$^{-2}$ at different temperatures: $T=0.2$, 0.8, 1.2 K. (b) The same data presented as
a function of $B^{2}$.\label{Eight}}
\end{center}
\end{figure}

In this structure we observe parabolic negative magnetoresistance (NMR), Fig. \ref{Eight}(a,b), which in
agreement with the prediction of interaction theory \cite{Gornyi} for a long-range random potential,
Eq.(\ref{eq5}). The $k_{F}d$ value varies from 1.2 to 2.2 which proves that the fluctuation potential with
the correlation length $d$ (spacer width) is indeed long-range (as the electron wavelength is smaller that
the correlation length of the potential). This is further supported by the fact that the momentum relaxation
time in these structures is much larger than the quantum lifetime ($\tau\gg\tau_{q}$). The magnetoresistance
is analysed in the range $\omega_c\tau>1$ to satisfy the condition of theory \cite{Gornyi}.

\begin{figure}[t]
\begin{center}
\includegraphics[width=0.6\textwidth]{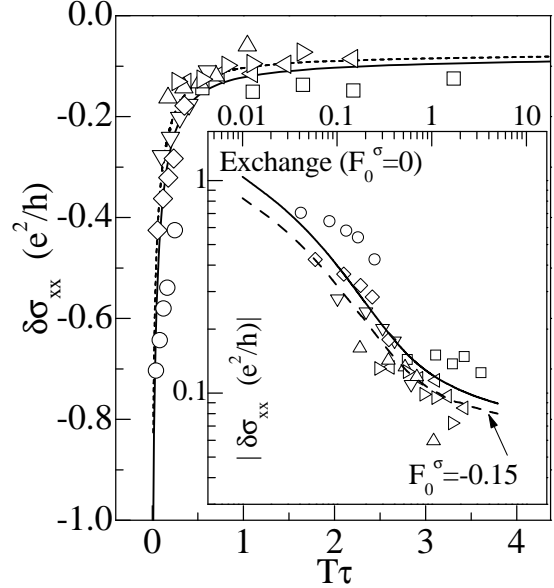}
\caption{Interaction correction obtained for different electron densities, $n=0.46-2.0 \times10^{11}$
cm$^{-2}$ (different symbols). Solid line - theoretical prediction \cite{Gornyi} for the correction due to
the exchange interaction, shifted by $-0.07e^{2}/h$. Dashed line - theory for the total correction with
$F_{0}^{\sigma}=-0.15$. Inset: the same results presented in the logarithmic scales.\label{Nine}}
\end{center}
\end{figure}

In Fig. \ref{Eight}(a) the NMR exhibits a sharp change in small fields caused by weak localisation, followed
by a parabolic dependence. We analyse the parabolic NMR in the range of fields well above the `transport'
magnetic field $B_{tr}=\hbar /4De\tau \sim 0.013$T in order to suppress weak localisation. (We have also
confirmed \cite{Lijun} that the magnetic field is not large enough for the development of the
magnetoresistance caused by the Zeeman effect on the interaction correction \cite{Altshuler}.)

In Fig. \ref{Eight}(b) the resistivity is plotted as a function of $B^2$, and from the slope of the straight
line $\delta \sigma _{xx}\left( T\right) $ is obtained. Fig. \ref{Nine} shows the temperature dependence of
$\delta \sigma _{xx}$ for different electron densities, where experimental points concentrate around one
curve. This curve becomes close to the interaction correction in the exchange channel \cite{Gornyi} if one
makes a vertical shift of the theoretical dependence by $\Delta\sigma=-0.07 e^2/h$ (there are no other
adjustable parameters). This shift means that there is a small contribution to the magnetoresistance which is
temperature independent. We believe \cite{Lijun} that the physical origin of this additional contribution is
the classical quadratic NMR \cite{Polyakov}.

It is important to emphasise that the comparison was made with the contribution from the exchange channel
only, however it is known that there is another (Hartree) term in interactions controlled by the parameter
$F_{0}^{\sigma}$. Comparing the total correction (exchange plus Hartree \cite{Gornyi}) with the experiment
shows that the Hartree contribution is much smaller than the exchange contribution. It can be seen in Fig.
\ref{Nine}(inset) that within experimental error the magnitude of the parameter $F_{0}^{\sigma}$ in our case
cannot be larger than $0.1-0.2$.

\section{Comparison of $F_0^\sigma \left( r_s\right) $ in different 2D systems}

A comparison of the parameter responsible for the carrier interaction contains a sensible assumption that the
character of carrier-carrier scattering can be similar in different structures. Such a comparison of
$F_0^\sigma$ obtained at different carrier densities is done as a function of $r_s$.

\begin{figure}[b]
\begin{center}
\includegraphics[width=0.6\textwidth]{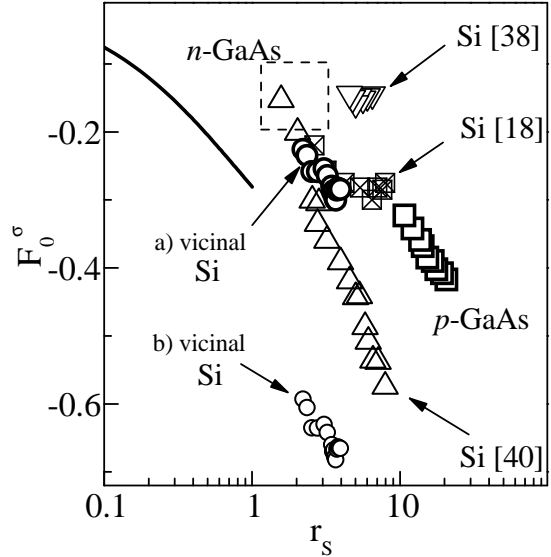}
\caption{The dependence of the Fermi liquid parameter on $r_{s}$ for different systems. The values of
$F_0^{\sigma}$ for the 2DEG in the vicinal Si structures have been found using the two-valley approach,
Eq.(\ref{eq7}), large circles, a), and one-valley approach, Eq. (\ref{eq2}), small circles, b). Dashed box
encloses the estimated values for the 2DEG in GaAs. Solid line is the theoretical curve for small $r_s$, from
\cite{Aleiner}. \label{Ten}}
\end{center}
\end{figure}
Firstly, let us compare the values of $F_{0}^{\sigma}$ obtained in our experiments on the 2DHG in GaAs and
2DEG in Si. One can see in Fig.\ref{Ten}(a) that the 2DHG results show a gradual decrease of the magnitude of
$F_0^\sigma$ with decreasing density. The obtained Si data nicely continue the trend of $F_0^\sigma(r_s)$
towards smaller $r_s$. However, if in the analysis of the results on Si we ignored the valley degeneracy
(section 4) and used the same expression for $\delta\sigma_{xx}(T)$ as for the 2DHG in section 3.2 (that is,
assuming strong intervalley scattering or large valley splitting $\Delta$), the obtained
$F_{0}^{\sigma}(r_s)$ would lie well below the general trend in Fig.\ref{Ten}(a). In this figure we also
indicate the range of the interaction parameter $F_{0}^{\sigma}$ obtained using theory \cite{Gornyi} on the
2DEG in GaAs with a long-range scattering potential. (Limited experimental accuracy does not allow us to
establish the density dependence $F_{0}^{\sigma}(r_s)$ in this case.)  It shows that our estimation of
$F_{0}^{\sigma}$ from the measurements of the 2DEG in GaAs is reasonable as it is consistent with other
results. The obtained trend in the dependence $F_{0}^{\sigma}(r_s)$ at $r_s\geq 1$ is also in agreement with
the calculations of such a dependence at $r_s\ll 1$ (the plotted relation is taken from \cite{Aleiner}).

Secondly, we compare our results with those recently obtained on the 2DEGs in $(100)$ Si MOSFETs
\cite{Shashkin,Vitkalov,Pudalov}. The authors also used the theory \cite{Aleiner} for point-like impurity
scattering. There is no full agreement here between different results, in spite of the measurements being
done on similar structures. This is partially caused by the fact that the analysis for the 2DEG in Si depends
strongly on the assumption about the  intensity of inter-valley scattering and the relation between the
valley splitting $\Delta$ and temperature. The difference in the assumptions could be reflected in the
determined value of $F_{0}^{\sigma}$, as different expressions are taken in the Hartree term of the
temperature dependence of the conductance. In \cite{Shashkin} it was assumed that $\Delta> k_BT$, while in
\cite{Vitkalov,Pudalov} the opposite assumption is made. The major discrepancy with our results is seen in
\cite{Vitkalov} where the obtained values of $F_0^\sigma$ are generally much smaller in magnitude, and in
\cite{Pudalov} where there is good agreement at smaller $r_s$ but the increase of the $F_0^\sigma$-value with
increasing $r_s$ appears to be much more rapid than in our work. These differences are interesting and
deserve further attention.

\begin{figure}[htbp]
\begin{center}
\includegraphics[width=0.95\textwidth]{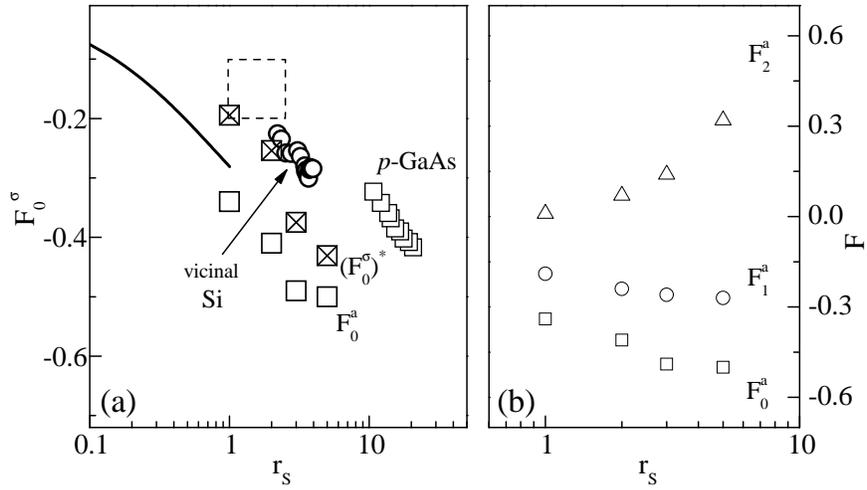}
\caption{(a) Comparison of the obtained values of the Fermi-liquid parameter with theoretical calculations.
Solid line is the dependence for $r_s\ll 1$ \cite{Aleiner}. The values of $F_0^a$ and ($F_0^\sigma)^*$ are
the results of calculations in \cite{Kwon}. (b) Comparison of the first three harmonics of the interaction
parameter $F^a$ from  \cite{Kwon}. \label{Eleven}}
\end{center}
\end{figure}

Finally, let us compare the obtained values of $F_{0}^{\sigma}(r_s)$ with theoretical calculations. As stated
earlier, the analytical expression for this dependence is only available at $r_s\ll 1$. For large $r_s$ there
are quantum Monte-Carlo calculations \cite{Kwon} which give the values of the Fermi-liquid parameters
$F_l^{s(a)}$ for several $r_s$-values. Fig.\ref{Eleven}(b) shows the comparison of our results with the
relevant parameter $F_0^{\sigma} \equiv F_0^{a}$ taken from \cite{Kwon}, which is clearly larger in magnitude
than the experimental Fermi-liquid parameter $F_0^{\sigma}$.
To comment on this disagreement one has to take into account the following feature of the interaction theory
\cite{Aleiner,AleinerB}. It contains an assumption that electron-electron interaction has a spherically
symmetrical character, so that only one constant, $F_0^\sigma$, is important. In general the interaction
function has an angular dependence and is represented as a sum of angular harmonics \cite{Kwon}:
\begin{equation}
F^{a(s)}(k,k^{'})=\sum_{l=0}^{\infty} F^{a(s)}_lcos(l\theta_{kk^{'}}).\label{eq9}
\end{equation}
The inset to Fig.\ref{Eleven}(b) shows the comparison of the first three harmonics $F_l^{a}$ ($l=0,1,2$)
calculated in \cite{Kwon}.  It is seen that the contribution of higher harmonics is not negligible compared
with $F_0^{a}$. Using from \cite{Kwon} the values of the three $F_l^{a}$-harmonics and also two
$F_l^{s}$-harmonics one can calculate the Hartree contribution and express it in terms of an effective,
averaged $(F_0^\sigma)^{*}$ \cite{Gornyi,Private} which can then be compared with experiment:
\begin{eqnarray}
\frac{3(F_0^\sigma)^{*}}{1+(F_0^\sigma)^{*}}&=&3\left(\frac{F_0^a}{1+F_0^a}-
\frac{F_1^a}{1+F_1^a}+\frac{F_2^a}{1+F_2^a}\right)\nonumber\\
&&-\frac{F_1^s}{1+F_1^s}+\frac{F_2^s}{1+F_2^s}. \label{eq10}
\end{eqnarray}
It is seen that taking into account extra harmonics makes better agreement with experiment,
Fig.\ref{Eleven}(a). This comparison cannot be totally conclusive as even higher-order harmonics should be
taken into account.  This result, however, can be treated as an indication that the reason for the observed
disagreement between the experimental $F_0^\sigma$ values and $F_0^a$ from \cite{Kwon} could be the angular
dependence of the interaction parameter.

\section{Conclusion}

We have demonstrated that the metallic character of $\rho \left( T\right) $ near the metal-to-insulator
transition and the positive magnetoresistance in parallel field of the 2DHG in GaAs structure with a
short-range random potential are determined by the hole-hole interaction in the ballistic limit, $k_BT\tau
/\hbar>1$. This conclusion is valid for hole densities $p=2-9\times 10^{10}$ cm$^{-2}$ where the value of the
Fermi-liquid parameter $F_0^\sigma \left( r_s\right) $, which determines the sign of $\rho \left( T\right) $,
has been obtained experimentally.

In zero magnetic field, the value of the interaction constant has also been obtained from the metallic $\rho
\left( T\right) $ of a 2DEG in vicinal Si MOSFETs, where electron-impurity scattering is also determined by a
short-range fluctuation potential.

The predictions of the interaction theory beyond the short-range approximation were tested in a 2DEG in a
GaAs heterostructure where the electron scattering is determined by a long-range fluctuation potential. In
strong magnetic fields we have observed a parabolic, temperature dependent negative magnetoresistance and
used it to find the electron-electron interaction correction in the intermediate and ballistic regimes. For
all studied 2D systems in the ballistic regime we have compared the obtained values of $F_0^\sigma(r_s)$.

\begin{acknowledgments}

We are grateful to I. L. Aleiner, B. N. Narozhny, I. V. Gornyi and A. D. Mirlin for discussions, EPSRC and
ORS award funds for financial support.
\end{acknowledgments}


\begin{chapthebibliography}{99}

\bibitem{Altshuler}  B. L. Altshuler and A. G. Aronov, in {\it
Electron-Electron Interaction in Disordered Systems}, edited by A. L. Efros and M. Pollak (North-Holland,
Amsterdam, 1985).

\bibitem{Finkelstein} A. M. Finkelstein, Sov. Phys. JETP \textbf{57}, 97 (1983).

\bibitem{Aleiner}  G. Zala, B. N. Narozhny, and I. L. Aleiner, Phys. Rev. B
{\bf 64}, 214204 (2001).

\bibitem{AleinerB}  G. Zala, B. N. Narozhny, and I. L. Aleiner, Phys. Rev. B {\bf 65}, 020201 (2001).

\bibitem{Abrahams} E. Abrahams, S. V. Kravchenko, and M. P. Sarachik, Rev. Mod. Phys.
{\bf 73}, 251 (2001).

\bibitem{Gangof4} E. Abrahams, P. W. Anderson, D. C. Licciardello, and T. V. Ramakrishnan, Phys. Rev. Lett. {\bf 42}, 673 (1979).

\bibitem{Ando} T. Ando, A. Fowler, and F. Stern, Rev. Mod. Phys. \textbf{54}, 437 (1982).

\bibitem{Shayegan} M. Shayegan, V. J. Goldman, C. Jiang, T. Sajoto, and M. Santos, Appl. Phys. Lett. {\bf 52},
1086 (1988); {\it ibid.} {\bf 53}, 2080 (1988).

\bibitem{Gold} A. Gold, Phys. Rev. B {\bf 41}, 8537 (1990); {\it ibid.} {\bf 44}, 8818 (1991).

\bibitem{Hirakawa86}  K. Hirakawa and H. Sakaki, Phys. Rev. B {\bf 33}, 8291
(1986).

\bibitem{Gornyi} I. V. Gornyi  and A. D. Mirlin, Phys. Rev. Lett. \textbf{90}, 076801 (2003).

\bibitem{Gornyi2} I. V. Gornyi  and A. D. Mirlin, cond-mat/0306029 (2003).

\bibitem{Paalanen}  M. A. Paalanen, D. C. Tsui, and J. C. M. Hwang, Phys. Rev. Lett. {\bf
51}, 2226 (1983).

\bibitem{Houghton}  A. Houghton, J. R. Senna, and S. C. Ying, Phys. Rev. B {\bf25,} 2196 (1982).

\bibitem{Proskuryakov} Y. Y. Proskuryakov, A. K. Savchenko, S. S. Safonov, M. Pepper, M. Y. Simmons, and D. A. Ritchie, Phys. Rev. Lett. \textbf{86}, 4895 (2001).

\bibitem{ProskuryakovInt} Y. Y. Proskuryakov, A. K. Savchenko, S. S. Safonov, M. Pepper, M. Y. Simmons, and D. A. Ritchie, Phys. Rev. Lett. \textbf{89}, 076406 (2002).

\bibitem{PudalovPRL} V. M. Pudalov, M. E. Gershenson, H. Kojima, N. Butch, E. M. Dizhur, G. Brunthaler, A. Prinz, and G. Bauer, Phys. Rev. Lett. \textbf{88}, 196404 (2002).

\bibitem{Shashkin} A. A. Shashkin, S. V. Kravchenko, V. T. Dolgopolov, and T. M. Klapwijk, Phys. Rev. B \textbf{66}, 073303 (2002).

\bibitem{ProskuryakovA} Y. Y. Proskuryakov, A. K. Savchenko, S. S. Safonov, L. Li, M. Pepper, M. Y. Simmons, D. A. Ritchie, E. H.
Linfield, and Z. D. Kvon, J. Phys. A \textbf{36}, 9249 (2003).

\bibitem{Stormer}  K. Hirakawa, Y. Zhao, M. B. Santos, M. Shayegan, and D. C. Tsui, Phys. Rev. B {\bf 47}, 4076
(1993); H. L. Stormer, Z. Schlesinger, A. Chang and D. C. Tsui, A. C. Gossard and W. Wiegmann, Phys. Rev.
Lett. {\bf 51}, 126 (1983).

\bibitem{VanHall} P. J. Van Hall, Superlatt. Microstruct. \textbf{6}, 213 (1989).

\bibitem{Karpus}  V. Karpus, Sem. Sci. Tech. {\bf 5}, 691 (1990).

\bibitem{Noh} H. Noh, M. P. Lilly, D. C. Tsui, J. A. Simmons, E. H. Hwang, S. Das Sarma, L. N. Pfeiffer, and K. W. West, Phys. Rev. B \textbf{68}, 165308 (2003).

\bibitem{DasSarmaHwa}  S. Das Sarma and E. H. Hwang, Phys. Rev. B {\bf 61}, R7838 (2000).

\bibitem{Mills}  A. P. Mills, A. P. Ramirez, L. N. Pfeiffer, and K. W. West, Phys. Rev. Lett. {\bf 83, }2805
(1999).

\bibitem{Murzin}  S. S. Murzin and S. I. Dorozhkin, JETP Lett. {\bf 67}, 113 (1998).

\bibitem{Sivan}  Y. Yaish, O. Prus, E. Buchstab, S. Shapira, G. Ben Yoseph, U. Sivan, and A. Stern, Phys. Rev. Lett. {\bf 84}, 4954
(2000).

\bibitem{Yaish01} Y. Yaish, O. Prus, E. Buchstab, G. Ben Yoseph, and U. Sivan, cond-mat/0109469 (2001).

\bibitem{gTutuc} E. Tutuc, E. P. De Poortere, S. J. Papadakis, and M. Shayegan, Phys. Rev. Lett. {\bf 86}, 2858
(2001).

\bibitem{Dolgopolov-Gold}  V. T. Dolgopolov and A. Gold, JETP Lett. {\bf 71}, 27 (2000).

\bibitem{DasSarma-parB}  S. Das Sarma, and E. H. Hwang, Phys. Rev. Lett. {\bf 84}, 5596 (2000).

\bibitem{mTutuc} E. Tutuc, S. Melinte, and M. Shayegan, Phys. Rev. Lett. {\bf 88},
036805 (2002).

\bibitem{ShashkinPRL} A. A. Shashkin, S. V. Kravchenko, V. T. Dolgopolov, and T. M. Klapwijk, Phys. Rev. Lett. {\bf 87}, 086801 (2001).

\bibitem{vanKesteren} H. W. van Kesteren, E. C. Cosman, W. A. J. A. van der Poel, and C. T. Foxon, Phys. Rev. B \textbf{41}, 5283 (1990).

\bibitem{Tutuc2} E. Tutuc, S. Melinte, E. P. De Poortere, and M. Shayegan, Phys. Rev. B \textbf{67}, 241309 (2003).

\bibitem{Zhu} J. Zhu, H. L. Stormer, L. N. Pfeiffer, K. W. Baldwin, and K. W. West, Phys. Rev. Lett. \textbf{90}, 056805 (2003).

\bibitem{Safonov}  S. S. Safonov, S. H. Roshko, A. K. Savchenko, A. G. Pogosov, and Z. D. Kvon, Phys. Rev. Lett. \textbf{86}, 272 (2001).

\bibitem{Vitkalov} S. A. Vitkalov, K. James, B. N. Narozhny, M. P. Sarachik, and T. M. Klapwijk, Phys. Rev. B. \textbf{67}, 113310 (2003).

\bibitem{PrivateNarozhny} B. N. Narozhny, private communication.

\bibitem{Pudalov} V. M. Pudalov, M. E. Gershenson, H. Kojima, G. Brunthaler, A. Prinz, and G. Bauer, Phys. Rev. Lett. \textbf{91}, 126403 (2003).

\bibitem{Lijun}L. Li, Y. Y. Proskuryakov, A. K. Savchenko, E. H. Linfield, and D. A. Ritchie, Phys. Rev. Lett. \textbf{90}, 076802 (2003).

\bibitem{Polyakov} A. D. Mirlin, D. G. Polyakov, F. Evers, and P. Wölfle,  Phys. Rev. Lett. {\bf 87}, 126805 (2001).

\bibitem{Kwon} Y. Kwon, D. M. Ceperley, and R. M. Martin, Phys. Rev. B \textbf{50}, 1684 (1994).

\bibitem{Private} I. L. Aleiner and B. N. Narozhny, private communication.

\end{chapthebibliography}
\end{document}